\documentclass[aps,prc,unsortedaddress,superscriptaddress,showpacs,twocolumn,floatfix]{revtex4}

\usepackage{amsmath}
\usepackage{graphicx}

\begin{document}

\title{Decay of negative muons bound in $^{27}$Al}

\affiliation{TRIUMF, Vancouver, BC, V6T 2A3, Canada}
\affiliation{University of British Columbia, Vancouver, BC, V6T 1Z1, Canada}
\affiliation{University of Montreal, Montreal, QC, H3C 3J7, Canada}
\affiliation{Texas A\&M University, College Station, TX 77843, U.S.A.}
\affiliation{Valparaiso University, Valparaiso, IN 46383, U.S.A.}
\affiliation{University of Regina, Regina, SK, S4S 0A2, Canada}
\affiliation{Kurchatov Institute, Moscow, 123182, Russia}

\author{A.~Grossheim}\email{alexg@triumf.ca}
\affiliation{TRIUMF, Vancouver, BC, V6T 2A3, Canada}

\author{R.~Bayes}
\altaffiliation[Affiliated with: ]{Univ.~of Victoria, Victoria, BC, Canada}
\affiliation{TRIUMF, Vancouver, BC, V6T 2A3, Canada}

\author{J.F.~Bueno}
\affiliation{University of British Columbia, Vancouver, BC, V6T 1Z1, Canada}

\author{P.~Depommier}
\affiliation{University of Montreal, Montreal, QC, H3C 3J7, Canada}

\author{W.~Faszer}
\affiliation{TRIUMF, Vancouver, BC, V6T 2A3, Canada}

\author{M.C.~Fujiwara}
\altaffiliation[Affiliated with: ]{Univ.~of Calgary, Calgary, AB, Canada}
\affiliation{TRIUMF, Vancouver, BC, V6T 2A3, Canada}

\author{C.A.~Gagliardi}
\affiliation{Texas A\&M University, College Station, TX 77843, U.S.A.}

\author{D.R.~Gill}
\affiliation{TRIUMF, Vancouver, BC, V6T 2A3, Canada}

\author{P.~Gumplinger}
\affiliation{TRIUMF, Vancouver, BC, V6T 2A3, Canada}

\author{M.D.~Hasinoff}
\affiliation{University of British Columbia, Vancouver, BC, V6T 1Z1, Canada}

\author{R.S.~Henderson}
\affiliation{TRIUMF, Vancouver, BC, V6T 2A3, Canada}

\author{A.~Hillairet}
\altaffiliation[Affiliated with: ]{Univ.~of Victoria, Victoria, BC, Canada}
\affiliation{TRIUMF, Vancouver, BC, V6T 2A3, Canada}

\author{J.~Hu}
\affiliation{TRIUMF, Vancouver, BC, V6T 2A3, Canada}

\author{D.D.~Koetke}
\affiliation{Valparaiso University, Valparaiso, IN 46383, U.S.A.}

\author{G.M.~Marshall}
\affiliation{TRIUMF, Vancouver, BC, V6T 2A3, Canada}

\author{E.L.~Mathie}
\affiliation{University of Regina, Regina, SK, S4S 0A2, Canada}

\author{R.E.~Mischke}
\affiliation{TRIUMF, Vancouver, BC, V6T 2A3, Canada}

\author{K.~Olchanski}
\affiliation{TRIUMF, Vancouver, BC, V6T 2A3, Canada}

\author{A.~Olin}
\altaffiliation[Affiliated with: ]{Univ.~of Victoria, Victoria, BC, Canada}
\affiliation{TRIUMF, Vancouver, BC, V6T 2A3, Canada}

\author{R.~Openshaw}
\affiliation{TRIUMF, Vancouver, BC, V6T 2A3, Canada}

\author{J.-M.~Poutissou}
\affiliation{TRIUMF, Vancouver, BC, V6T 2A3, Canada}

\author{R.~Poutissou}
\affiliation{TRIUMF, Vancouver, BC, V6T 2A3, Canada}

\author{V.~Selivanov}
\affiliation{Kurchatov Institute, Moscow, 123182, Russia}

\author{G.~Sheffer}
\affiliation{TRIUMF, Vancouver, BC, V6T 2A3, Canada}

\author{B.~Shin}
\altaffiliation[Affiliated with: ]{Univ.~of Saskatchewan, Saskatoon, SK, Canada}
\affiliation{TRIUMF, Vancouver, BC, V6T 2A3, Canada}

\author{T.D.S.~Stanislaus}
\affiliation{Valparaiso University, Valparaiso, IN 46383, U.S.A.}

\author{R.~Tacik}
\affiliation{University of Regina, Regina, SK, S4S 0A2, Canada}

\author{R.E.~Tribble}
\affiliation{Texas A\&M University, College Station, TX 77843, U.S.A.}

\collaboration{TWIST Collaboration}
\noaffiliation

\newcommand{\oalpha}{$\mathcal{O}(\alpha)$}

\date{\today}

\begin{abstract}
We present the first measurement of the energy spectrum up to 70 MeV of electrons
from the decay of negative muons after they become bound in $^{27}$Al atoms.
The data were taken with the TWIST apparatus at TRIUMF. We find a muon lifetime of 
(864.6 $\pm$ 1.2) ns, in agreement with earlier measurements. The asymmetry of the
decay spectrum is consistent with zero, indicating that the atomic capture has
completely depolarised the muons. The measured momentum spectrum is in reasonable
agreement with theoretical predictions at the higher energies, but differences around
the peak of the spectrum indicate the need for \oalpha~
radiative corrections to the 
calculations. The present measurement is the most precise measurement of the
decay spectrum of muons bound to any nucleus.
\end{abstract}


\maketitle

\newcommand{\e}[1]{\ensuremath{\times 10^{#1}}}
\newcommand{\us}{\ensuremath{\mu}s}      
\newcommand{\um}{\mbox{$\mu$m}}          
\newcommand{\kev}{keV}
\newcommand{\kevc}{keV/$c$}
\newcommand{\mev}{MeV}
\newcommand{\mevc}{MeV/$c$}
\newcommand{\mevcc}{MeV/$c^2$}
\newcommand{\pmuxi}{\ensuremath{P_\mu \xi}}
\newcommand{\pmupixi}{\ensuremath{P_\mu^\pi \xi}}
\newcommand{\abs}[1]{\ensuremath{\left\lvert#1\right\rvert}} 
\newcommand{\mudecay}{$\mu^- \rightarrow e^- \bar \nu_e \nu_\mu$}
\newcommand{\mumi}{$\mu^-$}
\newcommand{\muplus}{$\mu^+$}

\section{Introduction}

\label{sec:intro}

An elementary charged particle
can form an
atomic bound system when it replaces an atomic electron and/or a nucleus. Such
{\it exotic atoms} present interesting systems for both basic and applied
research, in topics ranging from quantum chemistry (e.g. pionic atom chemistry
\cite{Shinohara}), Coulomb three body systems (positronium ions \cite{Fleischer}), muon-catalysed nuclear
fusion (\cite{Fujiwara}), QED tests (Lamb shift \cite{PSI}), weak interactions (muon capture \cite{Andreev}),
strong interactions (hadronic shifts \cite{PRL}), as well as fundamental symmetry tests
(anti-hydrogen, antiprotonic helium \cite{Amoretti}).

The atomic structure of exotic atoms consisting of a heavy negative particle and a
nucleus has an unusual feature that, because of the larger mass $M$ of the negative
particle, its characteristic distance scale is smaller by $\sim m_e/M$ than
that of the ordinary atoms, and the Coulomb interactions are correspondingly larger.
The average potential energy and the particle velocity, respectively, are $V
\sim -(Z \alpha)^2 M$, and $\beta \sim Z \alpha$, where $Z$ is the nuclear
charge. Thus, these exotic atoms exhibit bound state effects that are
significantly more pronounced than in ordinary atoms.

A unique process that takes place in some classes of exotic atoms is
disintegration by decay of the short-lived constituent. 
The properties of such short-lived particles, such as the lifetime and the 
decay product energy spectrum,
are modified in the presence of the external fields of its binding partner.
Recently, universal bound state principles based on gauge symmetry have
attracted interest. These connect, for example, decay properties of
electromagnetically-bound exotic atomic states to those of heavy mesons,
quark-antiquark systems bound by the quantum chromodynamic gauge force \cite{Czarnecki1, Marciano, Pak}. 
The muonic atom is a system in which the decay properties can in principle
be calculated very precisely,
due to the purely leptonic nature of muon decay. Combined with the
strongly enhanced Coulomb interaction discussed above, it provides a sensitive
testing ground for our basic understanding of bound state modifications of
elementary processes.

Apart from its own interest as an exotic atomic system, there is currently
considerable interest in muonic aluminium atoms in the context of searches for
muon conversion to an electron. Two very ambitious proposals, Mu2e at Fermilab
\cite{mu2eproposal} and COMET at J-PARC \cite{cometproposal}, both propose 
to use muonic aluminium atoms to look for
extremely rare lepton flavor violating reactions. Electrons from muon decay in
the bound orbit of a muonic atom (DIO) near the end point are expected to give
the single largest source of background; hence they may limit the discovery potential
for these projects. 

In this paper, we report a measurement of the electron energy spectrum from DIO in
the muonic aluminium atom using the TWIST (TRIUMF Weak Interaction Symmetry Test)
spectrometer. Our result for the DIO spectrum is the first for aluminium and is 
the first precision measurement covering a considerable part of the spectrum in 
any element, dramatically
improving our experimental knowledge of DIO. We find that our measured spectrum is
not adequately described by the existing theoretical calculations, implying the
need for inclusion of radiative corrections. Our results will confront future
calculations of DIO, including an upcoming one based on the bound-state effective
field theory approach \cite{Czarnecki2,Czarnecki2p}.

Historically, the interest in the energy spectrum of the decay
electron for DIO has been focused on details of the nuclear field effects. It
was first calculated by Porter 
and Primakoff \cite{primakoff51}. They predicted a Doppler broadening of the 
spectrum by taking into account the momentum distribution of the muon in its 
bound state.
In subsequent papers \cite{ueberall60, gilinski60, huff61} the calculations were
expanded to include the modifications to the decay rate and its dependence on $Z$.
The spectrum was later recalculated with fewer approximations
\cite{haenggi74}, by taking into account higher order corrections to the 
nuclear potential \cite{fricke69}  and nuclear recoil. Furthermore, Herzog and
Alder \cite{herzog80} estimated the effects of bremsstrahlung in the nuclear 
field on the decay electron. Another calculation of the energy
spectrum and the asymmetry was done by Watanabe et al. \cite{watanabe87, watanabe93}
including tables of numerically calculated values for various elements.
While these calculations treat the muon-nuclear interaction in 
considerable detail, none of them have included the \oalpha~ 
radiative corrections that arise from the muon-electron interaction.

Despite the physics potential outlined by the authors of the calculations 
described above, to our knowledge, a sufficiently accurate measurement of the
energy spectrum does not exist.
Early measurements \cite{sargent55, block62, beilini68,suzukithesis} could confirm the 
expected features of Doppler broadening and the shift of the spectrum towards 
lower energies. However, these are by no means accurate enough to serve as 
tests for the calculations.
Later, data for a very limited portion of the spectrum were published for 
various elements in the context of $\mu-e$ conversion experiments. Due to 
the nature of those experiments, these results for Cu \cite{barthley64}, S 
\cite{sulfur81,sulfur82}, Ti \cite{triumf85, ahmad88, sindrum93}, Pb 
\cite{ahmad88,honeckersindrum06} and Au \cite{sindrum91, sindrum2006} focus on the 
high-energy tail of the distribution and can not easily be extrapolated 
towards lower energies due to systematic effects and normalisation problems. 
However, modern $\mu-e$ conversion experiments rely in their analysis on these 
calculated 
spectra (mostly \cite{herzog80} and \cite{watanabe93}) as input for simulations to 
estimate the expected background in their data.

The TWIST spectrometer was built for a high-precision experiment searching for
forms of the charged-current weak interaction in the decay of positive muons that
are not described by the Standard Model. To obtain a decay spectrum 
for a free, at-rest muon, $\mu^+$ are stopped in metal targets and the
angular and momentum spectra of decay positrons are measured.
\emph{Positive} (i.e. free) muons must be used for such measurements to avoid the
influence of the stopping target medium.

When a \emph{negative} muon comes to rest in matter, it is captured by
an atom and replaces an outer shell electron.
It then cascades almost instantly down to the $1s$ level by emitting X rays 
and Auger electrons. Two processes compete for the final fate of the muon: capture 
by the nucleus and DIO. The relative size of these two effects depends strongly on
the properties of the atom in which the muon is bound, and this can be determined by 
observing the electrons from the decay. In addition, for the muons that do decay 
before capture, the energy spectrum
of the resulting decay electrons
is modified from that of the decay of free muons: the energy spectrum is
shifted slightly towards lower energies as the electron must overcome the muon's
binding energy ($\approx 0.529$ MeV in Al). Also, the kinematic endpoint limit is 
$E_{max} \approx$ 105 MeV when the neutrinos carry away no momentum and the electron
recoils against the nucleus. These high-momentum decays are very rare, as close to 
the endpoint the spectrum drops sharply $\propto (E_{max} - E)^5$.

\section{Experimental Setup}
\label{sec:experiment}
A detailed description of the TWIST detector can be found in earlier
publications (see \cite{rob2008} and references therein). Here, an 
overview will be given with emphasis on components that are of particular 
interest for this analysis.

The TWIST spectrometer was located in the M13 beam line 
\cite{oram81:commis_m13_triumf} at TRIUMF. This channel provided a beam of negative 
muons created by a 500~MeV proton beam impinging on a Be target. The beam line was 
adjusted to {\it cloud muons} (muons generated in the proximity of the production
target) at a momentum of 29.6 \mevc~ and a rate of $\approx$ 80 Hz.
It had a considerable contamination with electrons that was largely eliminated
at the trigger level. Remaining electrons and a small amount of beam pions were
identified by scintillators in the beam line recording the time-of-flight (TOF)
and energy deposit of individual beam particles.

The muons were then transported into the centre of the detector and stopped
in a thin 71 $\mu$m aluminium target of 99.999\% purity where they decayed. The muon
range was adjusted using a gas degrader with variable density (adjustable ratio 
of He and CO$_2$), controlled by a feedback loop using the measured stopping 
position.

\begin{figure}[!hbt]
    \includegraphics[width=3.4in]{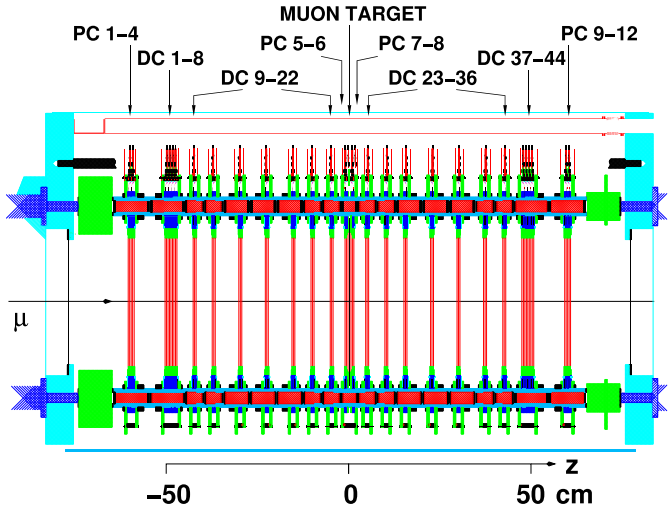}
    \caption{(Colour online.)  Side view of the 
      TWIST spectrometer planar chambers and support structures.
      Muons stopped in the target foil, which also served as a chamber
      cathode. The spectrometer is symmetric about the target foil, 
      immersed in a uniform 2.0 T solenoidal field aligned with 
      the beam axis.}
    \label{fig:halfstack}
\end{figure}

The tracks of decay electrons were measured with two symmetric stacks 
of 22 high-precision planar drift-chambers (DCs) \cite{TWIST_DCs:2005} 
located upstream and downstream of the target (Fig. \ref{fig:halfstack}).
In addition, a total of 8 multi-wire proportional chambers (PCs) were 
placed at the very upstream and downstream position of the detector to support
the event reconstruction by providing timing information. 
The target was surrounded by another 4 PCs (PC5/6 upstream and PC7/8
downstream) to enable the measurement of the stopping position of 
each individual muon. The target itself served
as the cathode foil of the two innermost PCs, thus the gas volumes surrounding
the target were sensitive. The complete detector was contained in a 
superconducting solenoid magnet, providing a highly uniform field of 2 Tesla.

A DC consisted of 80 sense wires at 4 mm pitch surrounded by Mylar cathode
foils of approximately 6 $\mu$m thickness and separated by 4 mm, filled with
dimethyl ether at atmospheric pressure. Most DCs were paired into modules of two 
(so-called $u$ and $v$ modules) with the central foil shared. A 
relative rotation of the wire planes by 90 degrees allowed for the 
reconstruction of the position of a hit in the perpendicular plane.
The PCs were of similar design, but their wire planes were equipped with 
twice as many wires and a ``faster'' gas (80:20 mixture of CF$_4$
and isobutane at atmospheric pressure) was chosen. In addition
to the timing information, the 
time-over-threshold was recorded to provide the amplitude of the signal, approximately
proportional to the energy deposit. All DC and PC wires, and scintillators were
read out by time-to-digital converters (LeCroy 1877 TDCs) in 0.5~ns bins from
6~\us\ before to 10~\us\ after a muon passed through the trigger scintillator. 
The space between the chambers was filled with a (97:3) mixture of helium and
nitrogen, with the relative pressure continuously being adjusted to avoid tension
and bulging of the chamber foils.
In total, there was approximately 140~mg/cm$^2$ of material from the vacuum of 
the M13 beam line through to the centre of the stopping target.

\section{Analysis}
\label{sec:analysis}

\subsection{Data Set}
\label{sec:dataset}
The data set considered for this analysis comprises 58M triggered events
containing 32M beam muons of which 5M stop in the target. For those, approximately
3M decays are observed inside the acceptance of the detector
of which around 1.3M remain after quality and kinematic fiducial cuts.

\subsection{Simulation}
\label{sec:simulation}
The TWIST detector is reproduced in a Monte Carlo (MC) simulation to permit the
study of the performance of the detector and the reconstruction, and to derive
corrections where necessary. The MC contains a detailed description of the detector
response and the physics processes affecting the decay tracks inside the detector.
This simulation is implemented in Geant~3.21 \cite{Geant321} and its correctness is
verified by direct comparisons between simulated and real \muplus~ data which is 
available with high statistics. In
particular the reconstruction efficiency, resolution, and bias and their phase-space
dependence have to agree. The complete procedure and results are described in more
detail in \cite{rob2008}. The output of the simulation is in the same format as the
real data and subject to the same calibration procedures (where applicable) and
reconstruction.

\subsection{Event Reconstruction}
\label{sec:reconstruction}
To reconstruct an event, the DC hits ---signal times on individual
wires--- are first grouped based on timing information from the PCs,
separately for the upstream and downstream halves of the detector.
A combinatorial, geometric pattern recognition is performed on the hits in such
a time window to assign groups of hits to a potential track candidate.
For each track candidate an initial helix fit is performed by using only the
spatial information of the hits, i.e. the crossing position of a pair of hit 
$u$ and $v$ wires in a module along with the module's $z$ position. This gives
the starting values for the drift-time fit (DTF). Here, the drift-time
information of each hit is used together with the space-to-time relation tables 
to position each hit inside the chamber volume. The DTF is not a simple
fit to a geometrical helix. In order to obtain better resolution and
minimise biases, the average energy loss of the particle and hence the diminishing
radius in the magnetic field is taken into account. Also, the deflections 
caused by multiple scattering are included in the fit by allowing kinks 
\cite{lutz88:kinks} at the centre of modules, weighted by the width of scattering
angles expected for the amount of crossed material. The DTF results in 
the particle's time, position and momentum vector at the chamber closest 
to the target that has contributed a hit. 
Therefore the track parameters describe the particle where it is first
seen, and not at the decay vertex. Although this difference
only has a small effect on the decay time, momentum and the projected angle of a 
track, it is systematic and is taken into account in the analysis.

Event and track selection cuts are then applied to remove background from the 
spectrum while maintaining a minimum of bias in the kinematic parameters of the
decay. A first set of selection criteria ensures that the beam particle is a muon
and that it stops in the target. The scintillators in the beam line
measure the TOF and pulse height per trigger, allowing
discrimination of muons from electrons and pions that are present in the beam.
The range of the muon is determined by requiring that the last hit
was recorded in PC6, directly in front of the target. In addition, pulse height
information of PC5 and PC6 is used to remove muons that stopped in
the gas and cause larger energy deposits. Finally, the position of the last two PC 
hits gives an estimate of the muon impact point, which should be within 3 cm of the 
detector axis. This ensures that the muon stops in the aluminium target, and 
not in the surrounding support material.
Decay electrons are then considered starting 500 ns after the arrival of
the muon. Earlier decays can suffer from reconstruction problems related to the 
overlap of the track ionisation with that of the
incoming muon, which would cause an upstream-downstream asymmetry.

The remaining events are decay candidates and the subsequent
track selections are aimed at removing potential background. Specifically,
decay tracks are required to have the correct charge sign, and their
extrapolated coordinates at the target must lie within close proximity of the
estimated muon impact point. In very rare cases, when more than one decay candidate 
is left at this stage, the one closer to the muon position is selected.

After these selections, kinematic fiducial cuts are applied in the $E-\cos\theta$
spectrum with $E$ being the total energy of the electron and $\theta$ the polar
angle with respect to the detector axis. These cuts serve to remove regions of the 
spectrum in which the reconstruction is known to be less reliable, or where the
simulation does not reproduce the data with sufficient accuracy. Specifically, 
limits are imposed on both $E$ and $\cos\theta$, as well as the transverse 
momentum $p_t$ (i.e. track radius) and longitudinal momentum $p_l$. The exact ranges
are adjusted depending on specific requirements of statistical and systematic
uncertainties as well as bias considerations. Kinematic fiducial cuts are applied
with respect to bin centres, not individual tracks.

For both data and simulation, the track reconstruction inefficiency within the
fiducial region is of the order of $10^{-3}$, the energy resolution is between 
50 and 150 \kev~ and the energy bias is less than 10 \kev.

\subsection{Spectrum Unfolding}
Due to the various error sources (biases, resolutions) and the limited acceptance
and efficiency of an experiment no measured observable represents the ``true''
physical value. The unfolding procedure tries to solve this problem and
to find the corresponding true distribution from a distribution of the measured
observable. The main assumption is that the probability distribution function
in the ``true'' physical parameters can be approximated by a histogram
with discrete bins. Then the relation between the vector $\vec{x}$ of the
true physical parameter and the vector $\vec{y}$ of the measured observable
can be described by a matrix $\mathcal{M}_{mig}$ which represents the mapping from the
true value to the measured one. This matrix, usually obtained from a simulation,
is called the migration (or response) matrix with
\begin{equation}
\vec{y}=\mathcal{M}_{mig} \cdot \vec{x} ~.
\label{eq:ufomig}
\end{equation}
In our case the $\vec{x}$ and $\vec{y}$ vectors contain the particle energy $E$ and
polar angle as $\cos\theta$.

The goal of the unfolding procedure is to determine a transformation
for the measurement to obtain the expected values for $\vec{x}$ using relation
\eqref{eq:ufomig}. The most simple and obvious solution is the inversion of the 
matrix. However, this method often provides unstable results. Correlations
between bins and statistical fluctuations cause the result to be 
dominated by large variances and strong negative correlations between neighbouring
bins.

In the method employed for this analysis \cite{ufo95}, the unfolding is performed
by the calculation of the unfolding matrix $\mathcal{M}^{ufo}$ in an iterative
way that is used instead of $\mathcal{M}_{mig}^{-1}$. Here $\mathcal{M}^{ufo}$
is a two-dimensional matrix that transforms a vector from the 
measurement space to the space of ``true'' values. As explained in detail in 
Ref. \cite{ufo95}, calculating the unfolding matrix this way avoids instabilities
while no assumptions on the migration matrix or the shape of the spectrum
have to be made. The method has been validated by unfolding simulated data.

To determine $\mathcal{M}_{mig}$, the MC is used to generate a sample of 
200M electrons with a decay spectrum flat in both $\cos\theta$ from -1 to 1 
and $E$ from 0 to 90 MeV. Since the statistical uncertainty of the MC
is propagated into the final result, the sample size is chosen large enough
to keep this contribution at an acceptable level.

\section{Results}
\label{sec:results}

\subsection{Lifetime}
The \mumi~ lifetime is obtained from a maximum likelihood exponential fit to the
track times from the helix reconstruction. As noted above, like all other 
track parameters, this time characterises the electron track at the first DC
encountered. In order to avoid large TOF corrections, all tracks that
do not have a hit in the first chamber adjacent to the target are
removed.
As this is typically caused by the pattern recognition, it does not
represent a time-dependent bias. This way only a small (sub-ns), angle-dependent 
correction needs to be applied to account for the approximate TOF
from the target to the first DC.

\begin{figure}[!hbt]
  \includegraphics[width=3.4in]{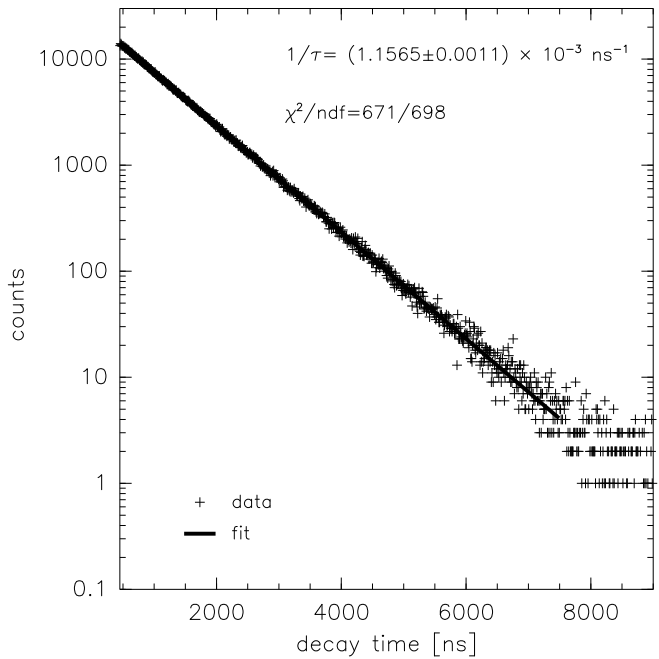}
  \caption{Reconstructed muon decay time spectrum. The decay time binning is 
10 ns, the fit range 500 to 7500 ns (error bars not shown).}
  \label{fig:decaytimespectrum}
\end{figure}

The measured decay time spectrum and the lifetime fit are shown in 
Fig. \ref{fig:decaytimespectrum}. The fit range is from 500 ns (below which
the reconstruction is not guaranteed to be time independent) to 7500 ns. In
addition to the uncertainties of the fit (0.8 ns), two sources of systematic 
uncertainties for this measurement are considered. Detector effects, including 
the non-linearity of the TDCs can be derived from a lifetime measurement of
free $\mu^+$, and contribute with an uncertainty 0.5 ns. Secondly, the muon selection
uncertainties using a variation on the PC5/6 cuts contribute with an
uncertainty of 0.7 ns.
We then find
\begin{equation}
\tau^{\text{Al}}_{\mu^-} = (864.6 \pm 0.8(\text{stat.}) \pm 0.9(\text{syst.})) ~
\text{ns}
\end{equation}
as the lifetime for our muon sample.

The observed decay time spectrum fits an exponential very well, and our result
is in excellent agreement with a published independent measurement \cite{suzuki87}.
This indicates that the selection of target stops using the PC 
amplitudes is of high purity.

\subsection{Asymmetry}
The angular asymmetry of the decay spectrum is directly proportional to the
polarisation $P_\mu$ of the muon. Thus, a measurement of the asymmetry can
be used to determine the muons' polarisation when the decay occurs. The TWIST 
apparatus and analysis have demonstrated the capability of measuring the
asymmetry of a decay spectrum with an accuracy of $1.7 \times 10^{-3}$ or better
in the $\mu^+$ analysis \cite{blair}.

Contrary to almost fully polarised {\it surface muons} (muons from pions at rest
that decay right at the edge of the target), cloud muons have a much lower initial
polarisation in the opposite direction. The exact treatment of the 
mechanisms that lead to a depolarisation while the muons travel to the target is 
not trivial. It is assumed that until they come to rest in matter, spin-changing 
interactions (mostly scattering processes) are independent of charge \cite{mudepol}. 
Thus, we can assume that the $\mu^-$ have a polarisation of about 
-0.25 \cite{rhopaper05} before the atomic capture.
The cascade following the capture will depolarise the muons further 
\cite{mudepol,spinrot}. The distribution of atomic states in which the capture 
occurs will
depend on the atomic structure and is not well calculated. Spin-flip transitions
during the cascade produce a large, nearly complete depolarisation in the
ground state.

A statistically significant difference in the number of decays
upstream and downstream is not observed. However, a quantitative limit on the
residual polarisation can be obtained from a fit of the complete angular
dependence of
the spectrum and a comparison with the same quantity for a $\mu^+$ data set.
With $P_{\mu^+} \approx -1$ and the polarisation 
depending linearly on the asymmetry, the ratio of asymmetries can be
used to calculate $P_{\mu^-}$.
Since the high-energy region 
of the spectrum is most sensitive to the asymmetry, a range of 31.5 to 52.5 \mev~
is used here.
In addition, a small modification for 
the bound $\mu^-$ decay has to be considered for this comparison; 
Ref \cite{watanabe93}
calculates that the expected integrated asymmetry should be $\approx7\%$ larger 
than for a free muon in the considered energy range.
After this correction we obtain
\begin{equation}
P_{\mu^-}=-0.005 \pm 0.003(\text{stat.}) \pm 0.0017(\text{syst.})
\end{equation}
as the residual 
polarisation when the muons decay after atomic capture.

\newcommand{\escale}{$2 \times 10^{-3}$}

\subsection{Energy Spectrum}
Finally, the decay energy spectrum is obtained from the unfolding procedure
described above. As can be expected from the high accuracy and efficiency
of the track reconstruction of the spectrometer, the unfolding corrects the
spectrum only marginally. Comparing the raw and corrected spectra,
the most significant difference is a small, angular-dependent shift of the 
energy scale. This is to be expected as the measured energy of each track
at the first DC has to be migrated to its energy at the decay vertex.

\begin{figure}[!hbt]
  \includegraphics[width=3.4in]{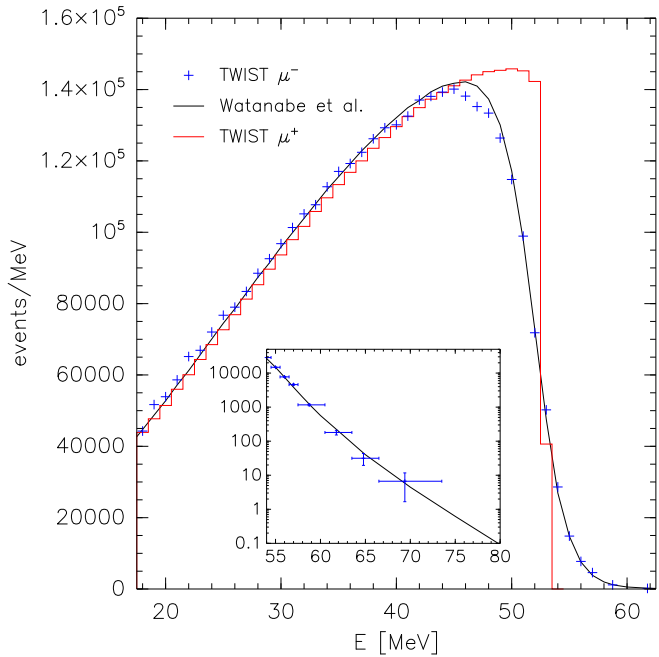}
  \caption{(Colour online.) Decay energy spectrum. For comparison, the appropriately normalised
    TWIST $\mu^+$ spectrum is shown as well as the spectrum calculated by
    Watanabe et al. \cite{watanabe93}. 
    Most error bars are smaller than the symbols and are not displayed. 
    The inset shows the high energy tail of the spectrum 
    on a logarithmic scale (\emph{with} error bars). The numerical values are
    tabulated in Table \ref{tab:momspec}. An overall energy scale error of \escale~
    is not included.}
  \label{fig:momspec}
\end{figure}

The unfolding is performed in the $E$-$\cos \theta$ space, but since no significant
angular dependence of the spectrum is observed ($P_{\mu^-} \simeq 0$ at the time of
decay), only projections are shown, including appropriate acceptance corrections.
The kinematic fiducial region of 0.54 $<|\cos\theta|<$ 0.92, $p_l>$ 14.0 \mevc,
11.0 \mevc~ $<p_t<38.0$ \mevc~ and 17.5 $\leq E <$ 73.5 MeV is used.
This is slightly different from the region chosen for previous analyses 
(e.g. \cite{rob2008}) in order to
optimise the stability of the results towards both the low and high energy tails of
the spectrum.
As a systematic uncertainty we assign an overall energy scale error of 
\escale (not included in Fig. \ref{fig:momspec} and Table 
\ref{tab:momspec}) to account for an uncorrected difference in the energy scale of
data and the simulation at the level of a few keV and the remaining systematic
uncertainties due to misalignments, differences in resolutions and biases between data
and MC.

In Fig. \ref{fig:momspec} the resulting energy spectrum is shown in comparison with
the theoretical spectrum calculated by Watanabe et al. \cite{watanabe93}. The 
relative normalisation is with respect to the number of observed \mumi decay 
events in the energy range of 17.5 to 73.5 MeV.
In the high-energy region ($\gtrsim$ 60 MeV) the spectrum decreases exponentially 
and the observed statistics are very small, requiring larger bin widths.
To account for this appropriately, the abscissa to which a bin is assigned is not the
bin centre, but given by the integral of the approximate exponential probability
density function of the population in that bin, normalised by the bin width itself
(see for example \cite{xlw}). This results in a shift of the energies of
the last four bins towards slightly smaller values.

\begin{figure}[!hbt]
  \includegraphics[width=3.4in]{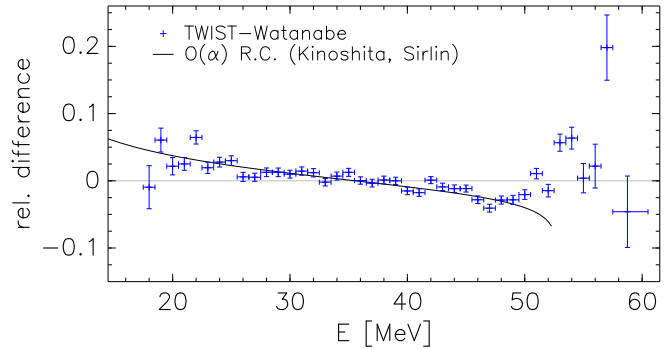}
  \caption{(Colour online.)
    Relative spectrum differences. The difference between the measured and calculated
    spectrum is normalised to the theoretical spectrum. The \oalpha~
    radiative corrections \cite{kinoshita} are normalised to the free muon decay
    spectrum with an endpoint of $\approx$ 52.8 MeV.
  }
  \label{fig:momspecdiff}
\end{figure}

We see a good agreement in the high energy tail, while in the peak region
there are significant differences between the spectra. At the same time, in the lower
end of the spectrum our measurement seems to be consistently above the prediction.
A comparison of these differences with the \oalpha~ radiative
corrections (R.C.) for the decay of free muons \cite{kinoshita} is shown in
Fig. \ref{fig:momspecdiff}. This demonstrates that in the region where the differences
between the decay spectra of free and bound muons are small ($\lesssim$ 46 MeV),
the free decay R.C. are applicable, as would be expected.
To what extent the observed differences above the free decay endpoint can be
attributed to approximations in Watanabe's spectrum, or the lack of radiative
corrections in the calculation will have to 
await dedicated calculations of such effects for the decay of bound muons.

\section{Summary}
\label{sec:summary}
We have measured the decay properties of muons bound in $^{27}$Al.
The excellent agreement of our measured lifetime with published data
demonstrates our
success in selecting a very pure sample of decays occurring within the thin aluminium
target material.
The depolarisation of the muons during capture must be complete as we find
negligible angular asymmetry in the decay spectrum.
According to our knowledge, the present measurement of the decay electron energy 
spectrum is the first 
to be performed on aluminium and the most accurate of all measured DIO 
spectra. Thus this 
is the first measurement that can be used to benchmark the calculated spectra
with precision.
In our comparison with the theoretical spectrum \cite{watanabe87} we do find
differences suggesting that \oalpha~
radiative corrections must be included before the 
assumptions about the basic physics
of exotic bound systems can be tested.

~\\
We would like to thank C.A.~Ballard, S.~Chan, B.~Evans, M.~Goyette,
and D.~Maas, D.~Mazur and the many undergraduate students who have
contributed to the construction, operation and analysis of TWIST. 
We also acknowledge many contributions by other staff members
from TRIUMF and collaborating institutes. A fruitful discussion with
W.~Marciano is also acknowledged.
This work was supported in part by 
the Natural Sciences and Engineering Research Council of Canada, the 
Russian Ministry of Science, and the U.S.A. Department of Energy. 
Computing resources for the analysis were provided by WestGrid.

\appendix
\section{Spectrum data}
\begin{table}
\begin{tabular}{|c|c|}
\hline
$E$ [MeV] & events/MeV \\
\hline
18 & $(4.43 \pm 0.14) \times 10^{4} $ \\
19 & $(5.169 \pm 0.086) \times 10^{4} $ \\
20 & $(5.391 \pm 0.067) \times 10^{4} $ \\
21 & $(5.863 \pm 0.054) \times 10^{4} $ \\
22 & $(6.517 \pm 0.060) \times 10^{4} $ \\
23 & $(6.692 \pm 0.054) \times 10^{4} $ \\
24 & $(7.203 \pm 0.051) \times 10^{4} $ \\
25 & $(7.674 \pm 0.053) \times 10^{4} $ \\
26 & $(7.900 \pm 0.053) \times 10^{4} $ \\
27 & $(8.341 \pm 0.050) \times 10^{4} $ \\
28 & $(8.851 \pm 0.056) \times 10^{4} $ \\
29 & $(9.258 \pm 0.058) \times 10^{4} $ \\
30 & $(9.680 \pm 0.055) \times 10^{4} $ \\
31 & $(1.013 \pm 0.0061) \times 10^{5} $ \\
32 & $(1.052 \pm 0.0064) \times 10^{5} $ \\
33 & $(1.077 \pm 0.0060) \times 10^{5} $ \\
34 & $(1.128 \pm 0.0066) \times 10^{5} $ \\
35 & $(1.171 \pm 0.0068) \times 10^{5} $ \\
36 & $(1.193 \pm 0.0065) \times 10^{5} $ \\
37 & $(1.224 \pm 0.0070) \times 10^{5} $ \\
38 & $(1.262 \pm 0.0072) \times 10^{5} $ \\
39 & $(1.293 \pm 0.0069) \times 10^{5} $ \\
40 & $(1.301 \pm 0.0073) \times 10^{5} $ \\
41 & $(1.326 \pm 0.0075) \times 10^{5} $ \\
42 & $(1.371 \pm 0.0072) \times 10^{5} $ \\
43 & $(1.381 \pm 0.0077) \times 10^{5} $ \\
44 & $(1.393 \pm 0.0078) \times 10^{5} $ \\
45 & $(1.401 \pm 0.0073) \times 10^{5} $ \\
46 & $(1.382 \pm 0.0079) \times 10^{5} $ \\
47 & $(1.352 \pm 0.0082) \times 10^{5} $ \\
48 & $(1.334 \pm 0.0082) \times 10^{5} $ \\
49 & $(1.264 \pm 0.0087) \times 10^{5} $ \\
50 & $(1.148 \pm 0.0084) \times 10^{5} $ \\
51 & $(9.892 \pm 0.075) \times 10^{4} $ \\
52 & $(7.184 \pm 0.068) \times 10^{4} $ \\
53 & $(5.022 \pm 0.060) \times 10^{4} $ \\
54 & $(2.861 \pm 0.043) \times 10^{4} $ \\
55 & $(1.484 \pm 0.032) \times 10^{4} $ \\
56 & $(7.738 \pm 0.24) \times 10^{3} $ \\
57 & $(4.599 \pm 0.18) \times 10^{3} $ \\
58.75 (59)& $(1.167 \pm 0.064) \times 10^{3} $ \\
61.77 (62)& $180 \pm 28$ \\
64.79 (65)& $32 \pm 12$ \\
69.39 (70)& $6.7 \pm 5.0 $ \\
\hline
\end{tabular}

\caption{TWIST $\mu^-$($^{27}$Al) decay-in-orbit energy spectrum. The energy
  coordinates for the last four bins are shifted and the bin centres are given in
  brackets (see text). 
  An overall energy scale error of \escale~ is not included.}
\label{tab:momspec}
\end{table}


\end{document}